\documentclass[prl,twocolumn,superscriptaddress]{revtex4}
\usepackage{amsmath,graphicx,bm,float}

\usepackage{color}

\newcommand{\ket}[1]{|{#1}\rangle}

\newcommand{\mean}[1]{\langle {#1}\rangle}

\begin{document}
\title{Experimental Test of Error-Disturbance Uncertainty Relations by Weak Measurement}
\author{Fumihiro Kaneda}
\altaffiliation{Present address: Department of Physics, University of Illinois, Urbana-Champaign, IL 61801}
\affiliation{Research Institute of Electrical Communication, Tohoku University, Sendai 980-8577, Japan}
\author{So-Young Baek}
\altaffiliation{Present address: Department of Electrical and Computer Engineering, Duke University, Durham, NC 27708}
\affiliation{Research Institute of Electrical Communication, Tohoku University,  Sendai 980-8577, Japan}
\author{Masanao Ozawa}
\affiliation{Graduate School of Information Science, Nagoya University, Nagoya 464-8601, Japan}
\author{Keiichi Edamatsu}
\affiliation{Research Institute of Electrical Communication, Tohoku University,  Sendai 980-8577, Japan}
\date{\today}

\begin{abstract}
We experimentally test the error-disturbance uncertainty relation (EDR) 
in generalized, strength-variable measurement of a single photon polarization qubit, 
making use of weak measurement that keeps the initial signal state practically unchanged. 
We demonstrate that Heisenberg's EDR is violated, 
yet Ozawa's and Branciard's EDRs are valid throughout the range of our measurement strength. 
\end{abstract}


\maketitle

The error-disturbance uncertainty relation (EDR) is one of the most fundamental issues in quantum mechanics since the EDR describes a peculiar limitation on measurements of quantum mechanical
observables.
In 1927, Heisenberg \cite{Heisenberg27} argued that any measurement of the position $Q$ of a particle with the error $\epsilon (Q)$ causes the disturbance $\eta (P)$ on its momentum $P$ so that
the product $\epsilon (Q)\eta(P)$ has a lower bound set by the Planck constant.
The generalized form of Heisenberg's EDR for an arbitrary pair of observables $A$ and $B$ is given by 
\begin{align}
 \epsilon (A) \eta (B) \geq C,
\label{HeisenbergEDR}
\end{align}
where $C = | \langle [ A,B ] \rangle |/2$, $[A,B] = AB - BA$, and $\langle ... \rangle$ stands for the mean value in a given state. 
It should be emphasized that Eq.  (1) is not equivalent to the following relation that is mathematically proven \cite{Kennard27, Robertson29}: 
\begin{align}
 \sigma (A) \sigma (B) \geq C,
\label{Robertson}
\end{align}
where $\sigma (A) = \sqrt{\langle A^2\rangle -\langle A \rangle^2}$ is the standard deviation. 
Indeed, Heisenberg's EDR  (1) is derived from (2) under certain additional assumptions
 \cite{AG88,Raymer94,Oza91,Ish91,Oza03a,Ozawa04},
but could fail where such assumptions are not satisfied.

In 2003, Ozawa \cite{Ozawa03} proposed an alternative EDR that is theoretically proven to be universally valid:
\begin{align}
\epsilon (A) \eta (B)+\epsilon (A) \sigma (B)+\sigma (A) \eta (B) \geq C . 
\label{OzawaEDR}
\end{align}
The presence of two additional terms indicates that the first Heisenberg's term $\epsilon (A) \eta (B)$ is allowed to be lower than  $C$, violating Eq.~(1).
To derive Eq.~(3),  
the error and disturbance were defined \cite{Ozawa03} 
for any general indirect measurement model depicted as a ``measurement apparatus (MA)'' in Fig. 1: 
\begin{align}
\epsilon (A) &\equiv \langle (U^{\dag}(I\otimes M)U- A\otimes I)^2\rangle^{\frac{1}{2}},  \notag \\
\eta (B) &\equiv \langle (U^{\dag}(B\otimes I)U- B\otimes I)^2\rangle^{\frac{1}{2}}, 
\label{eq.defed}
\end{align}
where  the average is taken in the state $\ket{\psi}_s \otimes \ket{\xi}_p$ of
the signal-probe composite system, 
$U$ is a unitary operator that provides interaction between the signal and probe systems, 
and $M$ is the meter observable in the probe to be directly observed.
The definition of $\epsilon(A)$ is uniquely derived from the classical notion of root-mean-square 
error if $U^{\dag}(I\otimes M)U$ and $A\otimes I$ commute \cite{Ozawa13}, 
and otherwise it is considered as a natural quantization of the notion of classical 
root-mean-square error.
The definition of $\eta(B)$ is derived analogously,  although there are recent debates 
on alternative approaches \cite{Ozawa13,Busch07,Watanabe11,Weston13,Busch13, Rozema13}.

Most recently, Branciard \cite{Branciard13} has improved Ozawa's EDR as
\begin{align}
 & \Bigl( \epsilon (A)^2 \sigma (B)^2  + \sigma (A)^2  \eta (B)^2 \Bigr. \notag \\
 & \Bigl. + 2  \epsilon (A) \eta (B) \sqrt{ \sigma (A)^2\sigma (B)^2-C^2} \Bigr)^{\frac12} \geq C,
\label{BranciardEDR}
\end{align}
which is universally valid and tighter than Ozawa's EDR.
Here $\epsilon (A)$ and $\eta(B)$ are still defined by Eq. (\ref{eq.defed}).
It is also pointed out  \cite{Branciard13} that the above relation becomes even stronger for spin measurements as described later.

\begin{tiny}
\begin{figure}[t!]
   \includegraphics[width=0.95\columnwidth,clip]{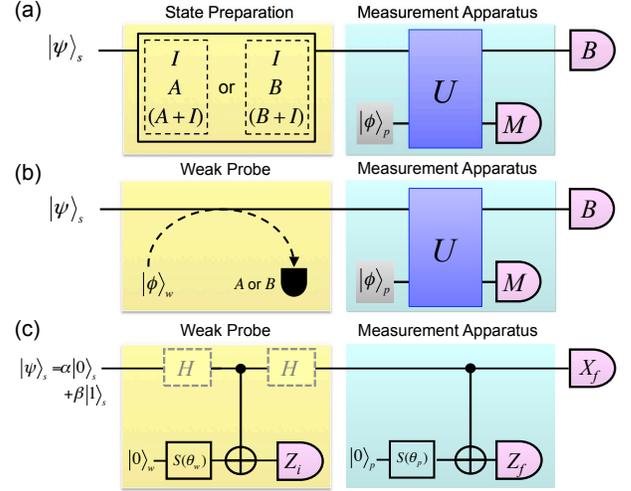}
   \label{two period qpm}
   \caption{Schematic diagram to test error-disturbance relation using (a) three state method and (b) weak-probe method. (c) Quantum circuit model of the weak-probe method for single-qubit observables $A = M = Z$ and $B = X$.   }
\end{figure}
\end{tiny}

For experimental test of EDRs, so far, two methods have been proposed: One is the so-called ``three-state method'' \cite{Ozawa04}, 
in which $\epsilon (A)$ for instance is obtained through the measurements 
of $M$ onto the prepared signal states,  
$\ket{\psi}_s$, $A\ket{\psi}_s$ and $(A + I)\ket{\psi}_s$, 
as shown in Fig.~1~(a).   
The three-state method was demonstrated in recent 
experimental tests of EDRs for quit systems:
projective measurement of a neutron-spin quibit \cite{Erhart12,Sulyok13} and generalized measurement of a photon-polarization qubit \cite{Baek13}. 
The other method is called ``weak-probe method'' \cite{Lund10, Ozawa05}.
In this method, as shown in Fig.~1~(b), 
a ``weak probe (WP)'' measures $A$ or $B$ with a weak measurement strength prior to the main measurement operated by MA.
When the measurement strength is sufficiently small, the signal state is sent to MA without disturbed by WP.
The three-state method is simpler to implement for a single qubit system, but 
the ``weak-probe method'' is more feasible in general case.

Lund and Wiseman \cite{Lund10}, and Ozawa \cite{Ozawa05} pointed out that the error (disturbance) defined by Eq.~(4) is given by root-mean-square difference between measurement outcomes of WP and MA (post-measurement of B):  
\begin{align}
\epsilon (A)^2 & = \sum_{i, f} (a_i-a_f)^2 P_{{\rm wv}} (a_i, a_f),  \notag \\
\eta (B)^2 & = \sum_{i, f} (b_i-b_f)^2 P_{{\rm wv}} (b_i, b_f),
\label{eq.ewv}
\end{align}
where 
$P_{{\rm wv}} (a_i, a_f)$ 
is the weak-valued joint probability distribution \cite{Steinberg95,Wiseman03}
taking  the outcomes $a_i$ in WP and $a_f$ in MA. 
As described later, we can experimentally estimate $P_{{\rm wv}} (a_i, a_f)$, 
and thus $\epsilon (A)$, 
by evaluating the probability distribution 
$P(a_i, a_f)$ that we take the outcomes $a_i$ and $a_f$. 
Similarly, $\eta (B)$ is  given by $P_{{\rm wv}}(b_i, b_f)$ taking outcomes $b_i$ in WP and $b_f$ in the post-measurement of $B$. 

Recently, Rozema {\it et al}.~\cite{Rozema12} experimentally demonstrated the experimental test of EDR for a single-photon polarization measurement using the weak-probe method.
They used a pair of entangled photons, one for a system qubit subjected to the main measurement and the other for an ancillary qubit subjected to the weak-probe measurement.
The state of the ancillary qubit after the weak-probe measurement was then ``teleported'' onto the system qubit and is subjected to the main measurement.
Although this fascinating scheme did work, in a real experiment it was rather complicated;
imperfect teleportation fidelity and rather strong measurement strength used for WP resulted in a considerable amount of disturbance on the system state.
As a consequence, the RHS of EDR was decreased to $C\sim0.8$ \cite{Rozema12} from its ideal value $C=1$.

In this letter, we report the experimental test of EDR for a single-photon polarization measurement using the weak-probe method. 
Our experiment uses only linear optical devices and single photons without entanglement, 
in a straightforward manner to the original proposal by Lund and Wiseman \cite{Lund10}.
Another advantage of our design is that it provides in principle no loss apparatuses for WP and MA,
unlike lossy apparatuses used in the previous experiment~\cite{Rozema12}. 
With this simple implementation, we can use sufficiently weak measurement strength for WP 
that causes very little disturbance on the signal state.
We show that our results clearly violate Heisenberg's EDR, yet validate both Ozawa's \cite{Ozawa03} and Branciard's \cite{Branciard13} relations.  
%

Our optical implementation of the weak-probe method is based on
the quantum circuit model \cite{Lund10} depicted in Fig. 1 (c). 
We take the signal observable to be measured as 
$A =  Z$ 
and $B = X$, where $X, Y$, and $Z$ denotes the Pauli matrices, 
and $\{ \ket{0}, \ket{1} \}$ 
are
the eigenbasis of $Z$ with the eigenvalues of 
$\{1, -1 \}$.  
The post-measurement observable for $X$ is $X_f$,
and the probe observable in MA and WP are $Z_f$ and $Z_w$, respectively.
Then, we use the following notation as the measurement outcomes: $a_{i,f} = z_{i,f} =\pm 1 $ and $b_{i,f} = x_{i,f} = \pm 1 $. 
We employ two cascaded circuits as WP and MA; both circuits work in the same manner.
In MA,
the probe qubit initialized to $\ket{0}_p$ is rotated by 
$S (\theta) = \begin{pmatrix} \cos \theta & \sin \theta \\ \sin \theta & -\cos \theta \end{pmatrix} $, 
where  $0 \leq \theta \leq \pi /4$. 
Then, the probe quibit is subjected to a controlled-NOT (CNOT) operation with the system qubit.
The POVM elements corresponding to the outcomes of $z_f = \pm 1$ are  \cite{Baek13}
\begin{align}
 E_{z_f=\pm1} = \frac{1}{2}(I \pm (\cos 2\theta) Z).
\label{eq.povm}
\end{align}
Here, $\cos 2\theta$ is the ``measurement strength'' of MA. 
By changing $\cos 2\theta$ from $0$ to  $1$, 
$E_{z_f=\pm1}$ change from identity (no measurement) to projector (strong measurement). 
WP works in exactly the same manner as MA 
except that the measurement strength of WP is $\cos 2 \theta_w$.  
In order to keep WP's measurement strength sufficiently weak,
$\theta_w$ should be close to $\pi/4$.
In addition, two Hadamard gates ($H$) are inserted to the system qubit before and after the CNOT in WP
when weak measurement for $X$ is taken. 
%


\begin{tiny}
\begin{figure}[t!]
  \includegraphics[width=0.95\columnwidth,clip ]{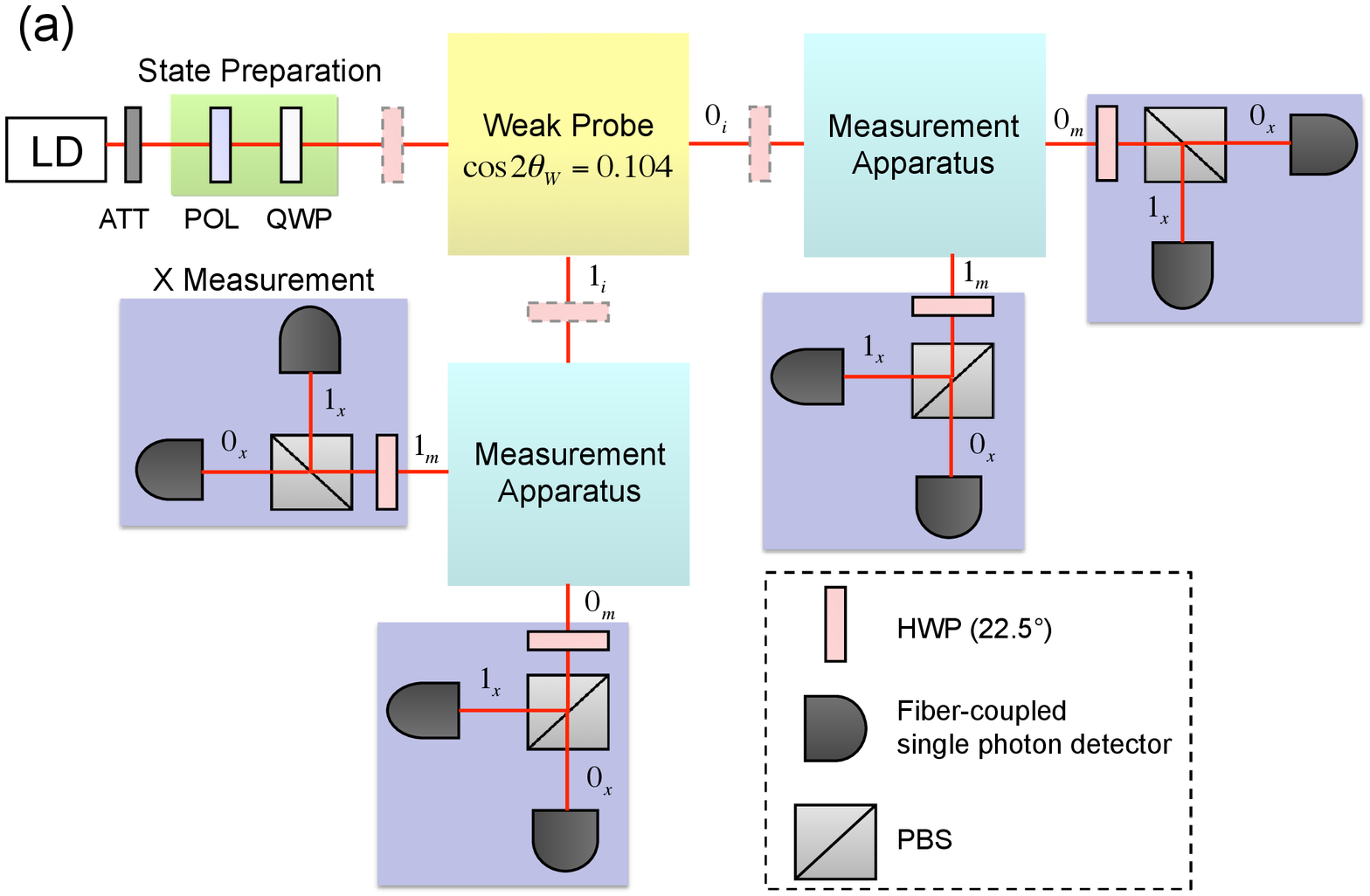}  
  \includegraphics[width=0.95\columnwidth,clip ]{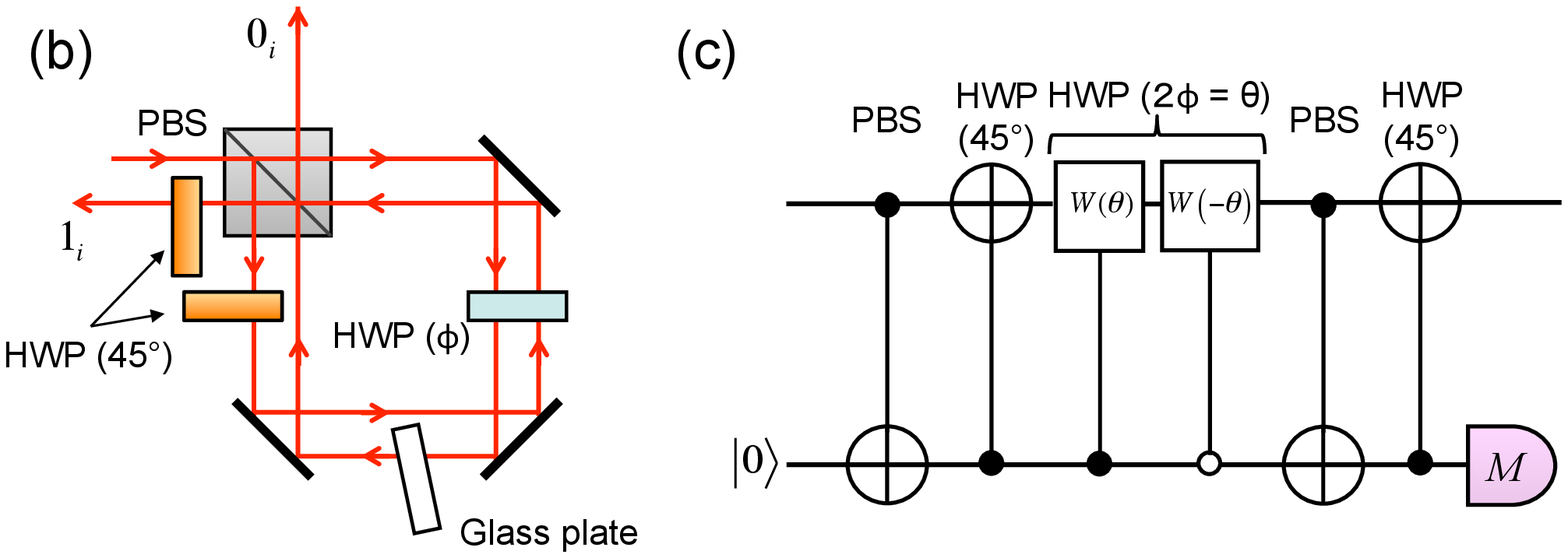} 
   \label{experimental setup}
   \caption{(a) Schematic diagram of experimental setup. Our optical implementation is clearly separated into the state preparation, weak probe, MA, and post $X$ measurement. Additional half-wave plates (HWPs, dashed rectangles) following and followed by the weak probe are only used for the investigation of $\eta (X)$. 
 (b) MA based on the Sagnac interferometer and (c) corresponding quantum circuit. The Sagnac interferometer is used for both weak probe and MA. The quantum circuit works as well as the one in Fig. 1 (c) when the initial probe state is $\ket{0}_p$.   }
\end{figure}
\end{tiny}

The experimental setup to test EDR by the weak-probe method is illustrated in Fig. 2 (a). 
In our experiment, horizontal and vertical polarizations,  
$\ket{H}$ and $\ket{V}$,  of a single photon
are chosen as the signal qubit with eigenstates $\ket{0}$ and $\ket{1}$ of $Z$, respectively.    
Thus, the measurement in MA corresponds to 
the polarization measurement in the $H-V$  basis
and does the post-measurement of $X$ to the $\pm45^\circ$ linear polarization basis.
Figure 2 (b) illustrates our optical implementation of 
WP and MA which are based on the idea of variable polarization beam splitter \cite{Baek08, Kim03,Baek13}. 
In the present experiment, 
we employed the displaced Sagnac configuration \cite{Nagata07}
that provide much higher phase stability than the Mach-Zehnder configuration used in our previous experiment \cite{Baek13}.
The corresponding quantum circuit of our instrument is shown in Fig.~2 (c),  
which provides the same POVM as that of Fig.~1(c) when the initial probe state is $\ket{0}_p$. 
Our probe qubit, initialized to $\ket{0}_p$, 
is encoded into the two path modes of the instrument
and the photon's output modes corresponds to the measurement outcome. 
For the WP and $X$ post-measurement, 
we use polarization beamsplitters (PBSs) with 
 $e_{r}\simeq100$ and $e_{t}>10^3$,
where $e_{r}$ and $e_{t}$ are the PBS's reflection extinction ratio and transmission extinction ratio \cite{Baek13}, respectively. 
For the PBSs used in MA,   
$e_{r}\simeq50$ and $e_{t}>10^3$.

As a single-photon source, we used a strongly attenuated continuous-wave diode laser (LD) whose center wavelength was at 686 nm, 
and the mean photon number existing in the whole apparatus at a time was $\sim$0.002.
To take the most stringent test of Ozawa's and Heisenberg's EDR, we chose the signal state as 
$\ket{\psi_0}_s = \left( \ket{H}+i \ket{V}\right)/\sqrt{2}$, 
an eigenstates of $Y$,
so that the RHS of the EDRs becomes the maximum value in the qubit measurement; 
$C = | \langle [ Z, X ] \rangle | / 2  = | \langle  Y \rangle | = 1$. 
We used
a polarizer (POL) and a quarter-wave plate (QWP) 
to prepare the signal qubit in 
$\ket{\psi_0}_s$. 
A half-wave plate (HWP) rotated at $22.5^{\circ}$ worked as a Hadamard gate for polarization qubits, rotating the photon's polarization by $45^\circ$.
The HWPs before and after WP changed the measurement basis of WP, between $Z$ and $X$.
In the experiment, the measurement strength of WP was set to $\cos 2 \theta_w = 0.104$ 
that produced very small disturbance in the initial signal state;
we expected $C = 0.995$, 
which was close to the ideal value $C=1$. 
Then, the signal photon was subjected to MA,
Because WP had two output outcomes,
we put two identical MAs after WP. 
At each output port of MA,  we put an instrument for the $X$ post-measurement,
consisting of a HWP, PBS and two photon counting detectors. 
We recorded the photon counting events $N_{ijk}$ in the single-photon detectors, where 
the subscript $i, j, k =0,1$ denotes the outcomes of the weak probe, MA, and $X$ post-measurement, respectively. 
From Eq.~(\ref{eq.ewv}) 
and the expression of  weak-valued joint probability distribution \cite{Lund10},
$\epsilon (Z)$  is given by
\begin{align}
\epsilon (Z)^2 
&= 2 \left( 1- \frac{1}{\cos\theta_w}\sum_{z_i,z_f} z_i z_f P(z_i, z_f) \right),
\label{eq.epd}
\end{align}
where $P (z_i, z_f)$ is the joint probability distribution taking the outcomes $z_i$ in WP and $z_f$ in MA.
Note that $\cos\theta_w$ is the measurement strength of WP.
$\eta (X)$ is  given by simply replacing $z_i$ and $z_f$ with $x_i$ and $x_f$, respectively 
To evaluate $\epsilon ( Z )$ and $\eta ( X )$ using Eq.~(\ref{eq.epd}), 
we experimentally obtain $P(z_i, z_f)$, and $P(x_i, x_f)$, 
analyzing the statistics of the single photon counting rates $N_{ijk}$ 
of the eight single-photon detectors. 
For instance, 
$P(z_i \!=\!1, z_f\!=\!1) = \sum_{k} N_{00k} / \sum_{i,k} N_{i0k}$.


\begin{tiny}
\begin{figure}[t!]
\includegraphics[width=0.95\columnwidth, clip]{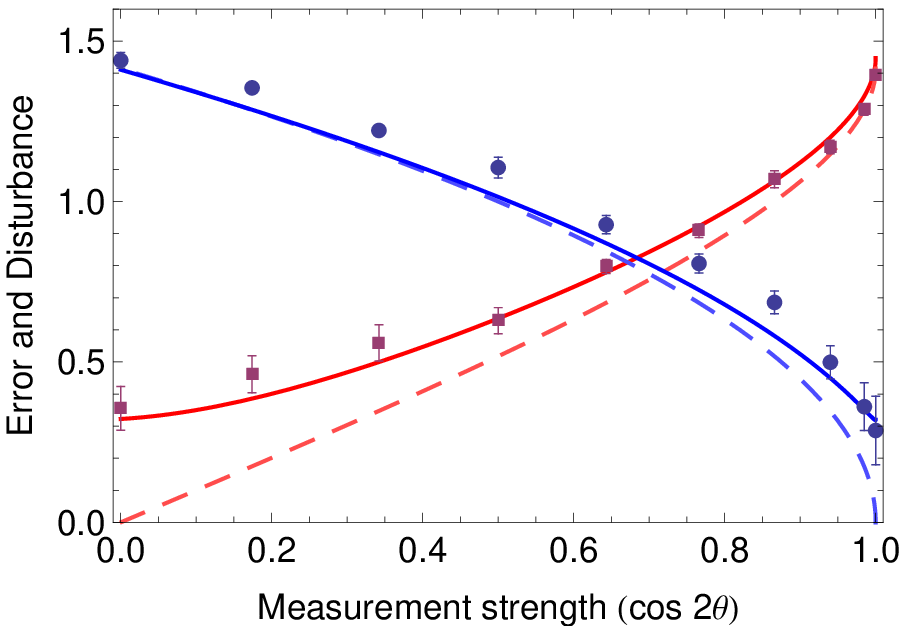}%
\vspace{3mm} \\
\includegraphics[width=0.95\columnwidth, clip]{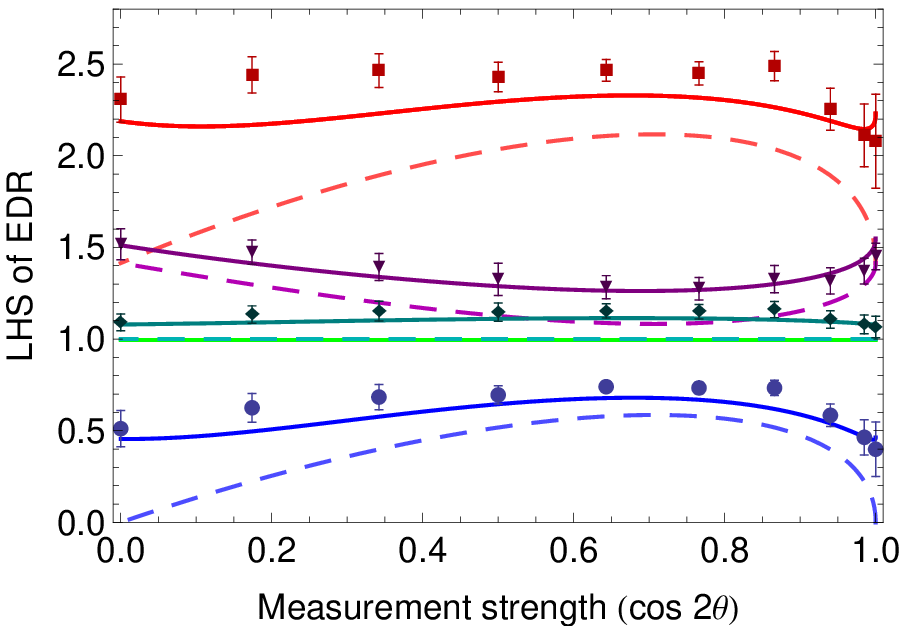}%
\caption{Experimental results. 
(a) the error $\epsilon (Z)$ (blue circles) and disturbance $\eta (X)$ (red squares) as functions of the measurement strength 
$\cos 2\theta$. 
Dashed curves are the theoretically calculated error and distuabance for perfect implementation of the quantum circuit presented in Fig. 1 (c). 
Solid curves are the theoretical values after the non-ideal extinction ratio of a PBS is taken into account. 
(b) Left-hand sides of the EDRs. Blue circles: Heisenberg's EDR in Eq. (\ref{HeisenbergEDR}). Red squares: Ozawa's EDR in Eq. (\ref{OzawaEDR}). Purple triangles: Branciard's EDR in Eq. (\ref{BranciardEDR}). Green diamonds:Branciard's EDR in Eq. (\ref{BranciardEDR2}). Dashed and solid curves are plotted in the same way as (a). 
}
\end{figure}
\end{tiny}

The quantities of $\epsilon (Z)$ and $\eta (X)$ thus obtained are shown in Fig. 3 (a). 
The error bars are obtained by RMS of repeated measurements for ten times. 
The dashed curves represent the theoretical calculations of  $\epsilon (Z)$ and $\eta (X)$ assuming the ideal instrument shown in Fig. 1 (c), and the solid curves are those in which the imperfect extinction ratio of the PBS taken into account (detailed discussion is given in Ref. \cite{Lund10, Baek13}). 
The experimentally measured error and disturbance present good agreement with the theoretical calculations. 
A small amount of systematic deviation from the calculation might originate from additional experimental imperfections 
that are not fully understood yet.
Nevertheless, 
we clearly see the trade-off relation between the error and disturbance; as the measurement strength increases, $\epsilon(Z)$ decreases while $\eta(X)$ increases. 
The experimental error and disturbance remain finite even when the other goes to zero in the ideal case, since the error and disturbance are given by RMS difference between $\pm1$-valued observables.  

From the experimentally measured error and disturbance, we evaluate the quantities of the LHS of the EDRs.  
We plot the LHS of 
Heisenberg's EDR (Eq.~(\ref{HeisenbergEDR}), bule),
Ozawa's EDR (Eq.~(\ref{OzawaEDR}), red),
and 
Branciard's EDR (Eq.~(\ref{BranciardEDR}), purple),
as shown in Fig.~3\,(b).
Also plotted is the stronger Branciard's EDR (green) that is applicable to the case (including ours) where the system and probe observables are both $\pm1$-valued and $\mean{A}$=$\mean{B}$=0 (hence $\sigma(A)$=$\sigma(B)$=1) \cite{Branciard13}
\begin{align}
 \left( \tilde\epsilon (A)^2  +  \tilde\eta (B)^2
 + 2  \tilde\epsilon (A) \tilde\eta (B) \sqrt{ 1-C^2}  \right)^{\frac12} \geq C, 
\label{BranciardEDR2}
\end{align}
where $\tilde\epsilon=\epsilon\sqrt{1-\epsilon^2/4}$ and $\tilde\eta=\eta\sqrt{1-\eta^2/4}$.
The solid and dashed curves are the theoretical predictions for each EDR with and without the imperfect extinction ratio of the PBS taken into account. 
In our experiment, the RHS of the EDRs is $C= 0.995$, which is indicated by the light green line.
Our experimental results demonstrate the clear violation of the Heisenberg's EDR, while the Ozawa's and Branciard's EDRs are always satisfied throughout the range of our measurement strength. 
We see that Branciard's EDRs are stronger than Ozawa's EDR; they are closer to the lower bound $C$ than Ozawa's.
In particular, LHS of Eq.~(\ref{BranciardEDR2}) saturates to the lower bound ($C=1$) for the ideal case.
It is also noteworthy that the experimental results are consistent with those reported in Ref. \cite{Baek13}, in which we used a similar apparatus and the three-state method to test Heisenberg's and Ozawa's EDRs.
%


\begin{tiny}
\begin{figure}[t!]
   \includegraphics[width=0.75\columnwidth, clip]{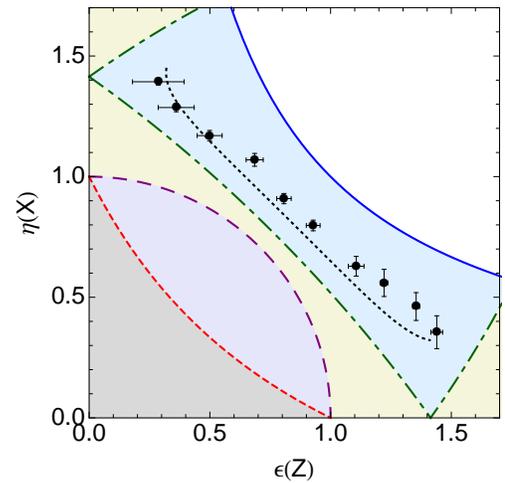} 
\caption{
Comparison of  EDRs' lower bounds in the error--disturbance plot.
Blue (solid) curve: Heisenberg's bound in Eq.~(\ref{HeisenbergEDR}). 
Red (short dashed) curve: Ozawa's bound in (\ref{OzawaEDR}).
Purple (long dashed) curve: Branciard's bound in (\ref{BranciardEDR}). 
Green (dot-dashed) curve: Branciard's bound in (\ref{BranciardEDR2}). 
Black (filled) circles: experimental data shown in Fig.~3~(a).
Black (dotted) curve: theoretical prediction for our experiment using imperfect PBSs.
The lower-left side of each bound is the forbidden region by the corresponding EDR.
Each bound was calculated for $C=1$.
}
\label{fig.ed2}
\end{figure}
\end{tiny}

In Fig.~\ref{fig.ed2}, we plot 
the predicted lower bounds of the EDRs in Eqs.~(\ref{HeisenbergEDR}), (\ref{OzawaEDR}), (\ref{BranciardEDR}) and (\ref{BranciardEDR2}), together with the experimental data.
Under Heisenberg' EDR the error or disturbance must be infinite when the other goes to zero,  while other EDRs allow finite error or disturbance even when the other is zero.
We again see that the experimental data violate Heisenberg's EDR, yet satisfy Ozawa's and Branciard's EDRs.
Our experimental data were close to Branciard's bound (dot-dashed curve) given in Eq.~(\ref{BranciardEDR2}),   
which could be saturated by ideal experiments.

In conclusion, we have experimentally tested the Heisenberg's, Ozawa's, and Branciard's EDRs 
in generalized photon polarization measurements making use of weak measurement
that keeps the initial signal state practically unchanged. 
Our experimental results clearly demonstrated that the Ozawa's and Branciard's EDRs were valid 
but Heisenberg's EDR was violated throughout the range of the measurement strength (from no measurement to projective measurement) . 
Such experimental investigation of the EDRs will be of demanded importance 
not only in understanding fundamentals of physical measurement 
but also in developing, for instance, novel measurement-based quantum information and communication protocols. 

While completing this manuscript, we became aware of a related work by M. Ringbauer {\it et al} \cite{Ringbauer13}.

The authors thank C. Branciard for valuable discussion.
This work was supported by MIC SCOPE No. 121806010 and the MEXT GCOE program. 

\end{document}